\documentclass[aps,pra,10pt,twocolumn]{revtex4-1}

\usepackage{amsfonts}
\usepackage{amsmath}
\usepackage{siunitx}
\usepackage{amssymb}
\usepackage{charter}
\usepackage{graphicx}
\usepackage{gensymb}
\usepackage{hyperref}
\UseRawInputEncoding
\begin{document}
	\title{Standard-quantum-limit-surpassing vector polarimetry using Rydberg atoms in an SU(1,1) interferometer}
	\author{Weiqiang Guan $^{1}$}
    \author{Yuetao Chen $^{1}$}
    \author{Jun Zhou $^{1}$}
    \author{Keyi Li $^{1}$}
    \author{Yu Huang $^{1}$}
	\author{Shaoyan Gao$^{1}$}
	\thanks{Corresponding author. gaosy@xjtu.edu.cn}
	\affiliation{$^{{\small 1}}$\textit{MOE Key Laboratory for Nonequilibrium Synthesis and
			Modulation of Condensed Matter, Shaanxi Province Key Laboratory of Quantum
			Information and Quantum Optoelectronic Devices, School of Physics, Xi'an
			Jiaotong University, 710049, China}
		}
	
	\begin{abstract}
	Vector polarimetry is an important application frontier for Rydberg-atom-based sensing. While prior research has largely concentrated on developing novel measurement schemes, high-sensitivity vector polarimetry remains an open question. Here we propose a theoretical framework for high-sensitivity detection of radio-frequency (RF) electric field polarization direction, which is particularly suitable for weak-field detection. Under a static magnetic field, the asymmetry in coupling between the Zeeman sublevels of the Rydberg atom and the RF field's polarization components enables the polarization angles to be determined from the atomic absorption index, which is retrieved via homodyne detection by incorporating the Rydberg atom system into an SU(1,1) interferometer. We derive the sensitivity of the polarization angles along with the corresponding standard quantum limit (SQL) and quantum Cram\'{e}r--Rao bound (QCRB). Our results demonstrate a sensitivity surpassing the SQL across wide angular ranges using either dual coherent states or a coherent state combined with a squeezed vacuum state as input. Significantly, the optimal sensitivity reaches below \SI{e-6}{\degree}, with sensitivities better than \SI{e-3}{\degree} maintained over most of the angular domain. This work establishes a foundation for high-precision vector polarimetry, thereby advancing the development of Rydberg-atom-based quantum sensing and contributing to a deeper understanding of light--matter interactions.	
	
	\end{abstract}
	
	\maketitle
	
	\section{Introduction}
	Rydberg atoms are pivotal for quantum sensing, benefiting from their large orbital radius, long lifetime, giant electric dipole moments, and rich energy-level structure \cite{gallaghar1994rydberg, Adams_2020}.The first theoretical and experimental demonstrations of Rydberg atom-based electrometers (RAEs) were achieved by three works \cite{5711293, 6910267, sedlacek2012microwave}. Following this initial breakthrough, RAEs began to advance rapidly while continuously broadening their applications, such as the measurement of phase \cite{10.1063/1.5088821, PhysRevApplied.17.044020, jia2021transfer, liu2022all} and orbital angular momentum (OAM) \cite{wang2025quadrupole,wang2026rydberg}. 
	
	Polarization, as a fundamental property of electromagnetic waves, characterizes their temporal evolution and is intrinsically linked to the wave's Direction of Arrival (DOA). Rydberg atoms offer a promising route to such high-sensitivity polarization measurement, which is crucial for applications like passive radar and electronic warfare that rely on precise DOA estimation of signal sources. Moreover, this technology also holds considerable promise for advancing quantum communication and quantum information processing.

	Several Rydberg-atom-based polarimetry (RAP) techniques have been developed, which can be broadly categorized into two categories: The first one introduces orthogonal polarized local (reference) fields into a Rydberg mixer, which generates beat signals with the signal electric field. The frequencies of these beat signals directly yield the relative polarization angles between the two fields. \cite{wang2023precise,yin2024measurement,elgee2024complete}. The second one exploits the asymmetric coupling strength for different polarization component relative to the quantization axis of the Rydberg atom to achieve polarization discrimination. For example, J. A. Sedlacek et al. employed a specific level configuration that prevented a subset of $\sigma$-polarized microwaves(MWs) from coupling to the high-lying Rydberg state to extract the polarization angles of the MW electric field \cite{sedlacek2013atom}. Yuechun Jiao et al. utilized the line-strength ratio between two Floquet levels with differing $m_j$ to extract the polarization angle \cite{jiao2017atom}. In the work of D. A. Anderson et al., a polarization-selective field enhancement resonator was used, which allowed coupling only for RF fields with a linear polarization component along the cavity's y-axis \cite{anderson2018vapor}. In the work of Chen et al., the determination of polarization angle was based on the relative areas of the Zeeman-resolved AT peaks \cite{chen2025polarization}. 
	
	Despite these advancements, the sensitivity of RAEs and RAP is ultimately limited by a finite signal strength in the presence of noise. The noise present includes laser power fluctuations, detector and electronic noise, photon shot noise (PSN), thermal radiation background fluctuations, and atom shot noise \cite{schlossberger2024rydberg, yuan2023quantum}. To overcome these limitations, several sensitivity-enhancement strategies have been developed, including: atomic superheterodyne detection \cite{jing2020atomic}, population repumping of ground state \cite{prajapati2021enhancement}, Rydberg criticality in a many-body atomic system \cite{ding2022enhanced}, noise-enhanced microwave sensor based on stochastic resonance\cite{wu2024nonlinearity}, cavity-enhanced light-matter interaction \cite{wang2023cavity, wang2025high} and laser-cooled atoms \cite{haitao2023approaching}. Recently, Shuhe Wu et al. reported combining RAEs with advanced quantum interferometrics to enhance MW field strength sensing, achieving an optimal sensitivity on the order of $\SI{e-11}{V/m/Hz^{1/2}}$, which beats the PSN  \cite{PhysRevApplied.20.064028, zhang2025theoretical}.
	
	Inspired by this interferometric method for RAEs, we propose and theoretically analyze a quantum interferometry scheme adapted for high-sensitivity vector polarization measurement, which is particularly suitable for weak-field detection. The scheme harnesses a $\Lambda$-type four-level Rydberg atom system dressed by a weak static magnetic field which lifts the degeneracy of the Zeeman sublevels and creates four possible transition pathways. The differential coupling strengths of these transitions to the RF field's polarization components map the polarization direction (defined by two polarization angles $\theta_a$ and $\theta_p$) onto the atomic absorption index. This mapping enables the high-sensitivity retrieval of the polarization direction by incorporating the Rydberg system into an SU(1,1) interferometer and employing homodyne detection for the atomic absorption index. Notably, because a simultaneous two-parameter estimation scheme is infeasible, it is necessary to adopt two sequential or independent single-parameter estimation. This requires rotating one or both polarization angles and performing at least two measurements to determine them.
	
	 We investigate two input configurations of the SU(1,1) interferometer: (i) dual coherent states and (ii) a coherent state combined with a squeezed vacuum state. We evaluate the polarization angle sensitivity and derive the corresponding standard quantum limit (SQL) and quantum Cramér-Rao bound (QCRB). We find that for input configuration (i), homodyne detection enables the sensitivity of polarization angles surpass the SQL within  an angular range of approximately $0.16\pi$. For configuration (ii), the sensitivity surpasses the SQL across a broader range of about $0.42\pi$. In both cases, the optimal sensitivity attains the \SI{e-6}{\degree} level, while sensitivities better than \SI{e-3}{\degree} are maintained over most of the angular domain.
	 
	 Furthermore, the influence of key parameters on the sensitivity has been investigated, and their optimal values are adopted in the presented calculations. 
	
	The paper is organized as follows. In Sec.~\ref{sec:section_ii}, we derive the input-output relation of the SU(1,1) interferometer and calculate the sensitivity based on homodyne detection. In Sec.~\ref{sec:section_iii}, we compute the SQL and the QCRB for polarization angle estimation. In Sec.~\ref{sec:section_iV}, we obtain the Hamiltonian describing the interaction between the $\Lambda$-type four-level Rydberg atom system and the RF electric field, and derives the resulting susceptibility. In Sec.~\ref{sec:section_V}, we presents numerical results demonstrating the determination of the two polarization angles via the absorption index, reports the sensitivity achieved via homodyne detection in comparison with the SQL and QCRB, and analyzes the influence of key parameters on the sensitivity.
	
	\section{SENSITIVITY OF ABSORPTIVE MEASUREMENT WITH DISPERSION IN SU(1,1) INTERFEROMETER}\label{sec:section_ii}
	The probe mode of the SU(1,1) interferometer also couples to the ground-state transition of the Rydberg atoms. Therefore, before introducing the RAP, we first establish a model for the absorptive measurement in an SU(1,1) interferometer in which the Rydberg atom system is modeled as a phase change followed by a fictitious beam splitter.
	\begin{figure}[tbp]
		\includegraphics[width=1\columnwidth]{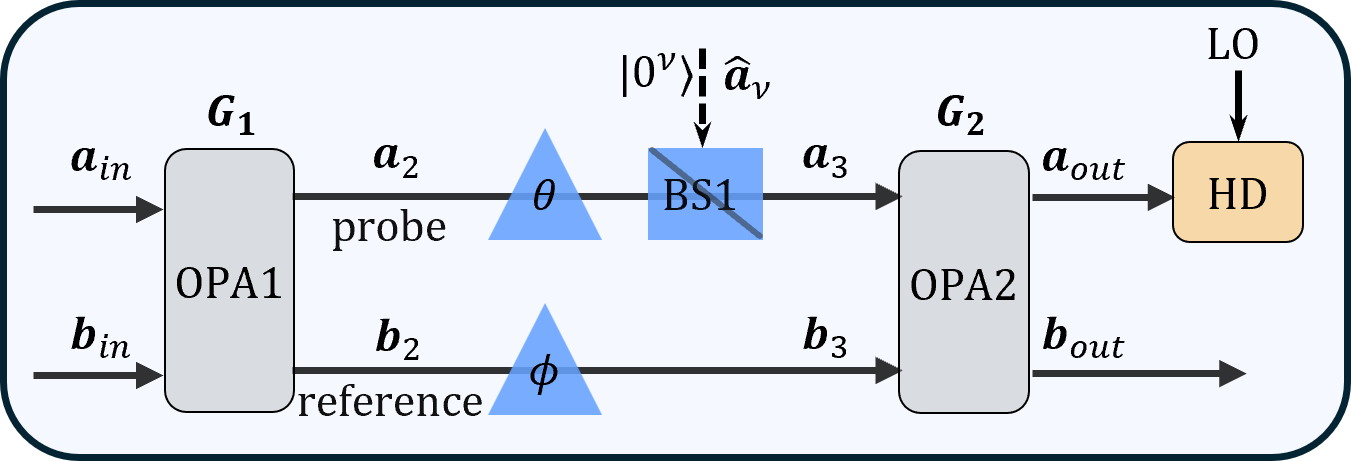}%
		\newline
		\caption{Schematic of absorptive measurement with dispersion in an SU(1,1) interferometer. The phase change $\theta$ is introduced prior to the fictitious beam splitter BS1, while phase shifter $\phi$ is used to balance the SU(1,1) interferometer. The state $|0^{\nu}\rangle$ and the operator $\hat{a}_{\nu}$ are vacuum field and its corresponding annihilation operator induced by the absorption measurement. OPA: optical parametric amplifier; HD: homodyne detection; LO: local oscillator.}
		\label{fig:1} \centering
	\end{figure}
	\subsection{Input-output relation of the SU(1,1) interferometer}\label{sec:subsection_ii_A}
	A schematic of the absorptive measurement with dispersion in SU(1,1) interferometer is shown in Fig.~\ref{fig:1}. We consider two input modes, denoted by the annihilation operators $\hat{a}_{in} $ and $\hat{b}_{in} $, which can be prepared either as coherent states or squeezed vacuum states. The first optical parametric amplifier (OPA1) is described by the unitary operator $\hat{S}(\xi _{1})=\exp (\xi _{1}^{\ast }\hat{a}_{in}\hat{b}_{in}- \xi_{1}\hat{a}_{in}^{\dagger}\hat{b}_{in}^{\dagger})$ where $\xi_{1}=g_{1}e^{i\theta _{1}}$ denotes the complex gain of OPA1. After going through OPA1, one of two outputs $\hat{a}_2$ serves as the probe mode and interacts with the Rydberg atomic ensemble within the vapor cell. The atomic medium is modeled as a phase change $\theta$ and a fictitious beam splitter (BS1) with transmissivity $T_1=e^{-\epsilon}$, where the $\epsilon$ is absorption index. This combined atomic interaction is formally represented by a sequence of two unitary operations:  $\hat{U}_{\theta}(\epsilon)=\exp (i\theta\hat{a}^{\dagger}_2\hat{a}_2)$ ,followed by	$\hat{U}_{1}(\epsilon)=\exp [\arccos(T_1)(\hat{a}_{\nu}\hat{a}_{2}^{\dagger}-\hat{a}_{\nu}^{\dagger}\hat{a}_{2} )]$, where $\hat{a}_{\nu}$ is the vacuum field induced by the absorptive measurement at BS1 and $\hat{a}_{2}$ is the mode entering the atomic ensemble. It is crucial to note that the order of these operations ($\hat{U}_{\theta}$ followed by $\hat{U}_{1}$) is physically fixed and cannot be commuted.​ Swapping the order would result in $\hat{U}_{\theta}$ acting on  $\hat{U}_{1}^{\dagger}\hat{a}_2\hat{U}_{1}$ rather than on $\hat{a}_2$ leading to inequivalent outcomes. In parallel, another mode $\hat{b}_2 $ serves as a reference. It passes through a tunable phase shifter $\phi$ to balance the phase change caused by the Rydberg atomic ensemble, ensuring that the detected signal reflects only the amplitude change due to absorption of Rydberg atom. Subsequently, the two output modes recombine in the second optical parametric amplifier (OPA2). By calculating the full input-output relationship of the SU(1,1) interferometer, accounting for all components, the final output field is derived as:

		\begin{equation}
			\hat{a}_{out}=W_1\hat{a}-W_2\hat{b}^{\dagger}+W_3\hat{a}_{\nu},
		\end{equation}
	with
	\begin{small} 
		\begin{equation}
			\begin{split}
				W_1 &= e^{i(\theta-\phi)}T_1\cosh{g_1}\cosh{g_2}+e^{i(\theta_2-\theta_1)}\sinh{g_1}\sinh{g_2}, \\
				W_2 &= e^{i\theta_1}T_1\sinh{g_1}\cosh{g_2}+e^{i\theta_2}\cosh{g_1}\sinh{g_2},\\
				W_3 &= \sqrt{1-T_1^2}\cosh{g_2},
			\end{split}
		\end{equation}
	\end{small}
	where $g_i$ and $\theta_i$ ($i=1,2$) are the gain factor and phase shifts of the OPA$_i$, respectively. In the following calculation, we set $\theta_2=\pi$, $\theta_1=0$, $g_1=g_2=g$ for simplicity and $\phi=\theta$ to balance the phase change caused by the Rydberg atomic ensemble.

	\subsection{Sensitivity of polarization via homodyne detection} \label{sec:subsection_ii_B}
	To investigate the polarization sensitivity, a local coherent mode (LO) and the output mode undergo homodyne detection, with the corresponding detected variable being the amplitude quadrature $\hat{X}_i=(\hat{a}_{out}+\hat{a}_{out}^{\dagger})/\sqrt{2}$, where $i=c,q$ denotes the case of dual coherent states or a coherent state combined​ with a squeezed vacuum state as the input, respectively. The slope (signal derivative) and noise (variance) of the amplitude quadrature $\hat{X_i}$ will jointly limit the sensitivity of polarization angles , which can be given by \cite{guo2020distributed,giovannetti2011advances}
	\begin{small} 
		\begin{equation}
			\Delta\theta_{ij}=\frac{\sqrt{\Delta^2\hat{X}_i}}{\vert\partial\langle\hat{X}_i\rangle/\partial\epsilon\vert}\times\frac{1}{\left\vert\partial\epsilon/\partial\theta_j\right\vert},\label{eq:3}
		\end{equation}
	\end{small}
	where $\theta_j$ ($j=a,p$) denotes the two polarization angles, $\Delta\theta_{ij}$ denotes the sensitivity for estimating $\theta_j$ under the input configuration $i$. Here, the absorption index $\epsilon$ serves as an intermediate variable in the error propagation. While the term $|\partial\epsilon/\partial\theta_j|^{-1}$ is evaluated in the RAP model, we now focus on analyzing the term ${\sqrt{\Delta^2\hat{X}_i}} / {|\partial\langle\hat{X}_i\rangle/\partial\epsilon|}$.
	
	We first consider the configuration with dual coherent states as inputs. The input mode $\hat{a}_{\mathrm{in}}$ is $|\alpha\rangle$ with $\alpha = |\alpha| e^{i\theta_{\alpha}}$, and the input mode $\hat{b}_{\mathrm{in}}$ is $|\beta\rangle$ with $\beta = |\beta| e^{i\theta_b}$. The slope and the noise of the observable $\hat{X}_c$ are then derived as
	\begin{small} 
		\begin{equation}
			\begin{split}
				\frac{\partial\langle\hat{X}_c\rangle}{\partial\epsilon} &=\frac{1}{\sqrt{2}}[(W_1^{'}\alpha+W_1^{'\ast}\alpha^{\ast})-(W_2^{'}\beta^{\ast}+W_2^{'\ast}\beta)],\\
				\Delta^2\hat{X}_c &= \frac{1}{2}\left(\left|W_1\right|^2+\left|W_2\right|^2+\left|W_3\right|^2\right),
			\end{split}
		\end{equation}
	\end{small}
	where $W_1' = \partial W_1/\partial\epsilon = -T_1\cosh^2{g}$, $W_2' = \partial W_2/\partial\epsilon = -T_1\sinh{g}\cosh{g}$, and $\langle\hat{X}_c\rangle$ and $\Delta^2\hat{X}_c$ denote the expectation value and variance of $\hat{X}_c$ for the given input states.
	
	Next, we consider the configuration with a coherent state combined with a squeezed vacuum state. The input mode $\hat{a}_{in}$ is $\vert\alpha\rangle$ with $\alpha = |\alpha| e^{i\theta_a}$, while the input mode $\hat{b}_{in}$ is $\hat{S}(\xi)\vert 0 \rangle$ with $\xi=re^{i\theta_r}$. Then $\hat{b}_{in}$ can be expressed as 
		\begin{equation}
			\hat{b}_q=\cosh {(r)}\hat{b}_0-\sinh {(r)}\hat{b}_0^{\dagger} e^{i\theta_r},
		\end{equation}
	 where $r$ and $\theta_r$ are the amplitude and angle of the squeeze parameter, $\hat{b}_0$ is the vacuum state. Similarly, we obtain the slope and the noise of the observable $\hat{X}_q$ as
	 \begin{small} 
	 	\begin{equation}
	 		\begin{split}
	 			\frac{\partial\langle\hat{X}_q\rangle}{\partial\epsilon} &=\frac{1}{\sqrt{2}}(W_1^{'}\alpha+W_1^{'\ast}\alpha^{\ast}),\\
	 			\Delta^2\hat{X}_q &= \frac{1}{2}( \left|W_1\right|^2+\left|\widetilde{W}_2\right|^2+\left|W_3\right|^2),
	 		\end{split}
	 	\end{equation}
	 \end{small} 	 
	where $\widetilde{W}_2 = W_2\sinh{(r)}e^{-i\theta_r}-W_2^{\ast}\cosh{(r)} $. In both configurations, the coherent amplitude $|\alpha|$ influences only the expectation value of the amplitude quadrature $\hat{X}$, and a larger amplitude theoretically enhances the sensitivity. However, increased laser power also raises collision rates and power broadening, making the choice of amplitude critical. Similarly, the amplitude $r$ affects the variance $\Delta^2\hat{X}_q$ and thus the sensitivity. A detailed analysis of these parametric influences is presented in Section~\ref{sec:subsection_V_C}
	
\section{SQL AND QCRB OF POLARIZATION ANGLES}\label{sec:section_iii}
In this section, two sensitivity bounds, SQL and QCRB are introduced to evaluate our quantum-enhanced polarization measurement scheme. The SQL represents the ultimate sensitivity achievable within a classical measurement framework,, while the QCRB is the ultimate lower bound for quantum measurements. To obtain the sensitivity bounds of the polarization angles, the quantum fisher information (QFI) of the probe state is calculated. In the lossless case and for a pure quantum state, the QFI for estimating a polarization angle $\theta_j$ can be expressed as \cite{liu2014quantum}
\begin{small} 
	\begin{equation}
		F_j(\theta_j) = 4\left[ \langle\psi_j'|\psi_j'\rangle - \left|\langle\psi_j'|\psi\rangle\right|^2 \right],
	\end{equation}
\end{small}
where $\vert\psi(\theta_j)\rangle = \hat{U}_1(\epsilon(\theta_j))\hat{S}(\xi_1)\vert\psi^{in}\rangle\vert0^{\nu}\rangle$ is the state vector after the BS1, $\vert\psi^{in}\rangle$ represents the input state and $\vert0^{\nu}\rangle$ is the vacuum state induced by absorptive measurement at BS1. $\vert\psi_j'\rangle = {\partial\vert\psi\rangle}/{\partial\theta_j}= {\partial_{\theta_j}\vert\psi\rangle}$. Then for our scheme, the QFI of polarization angles can be expressed as
\begin{small} 
	\begin{equation}	
		F_j(\theta_j) = 4\left[\langle\hat{H}_j^2\rangle-\langle\hat{H}_j\rangle^2\right],
	\end{equation}
\end{small}
where $\hat{H}_j = i{(\partial_{\theta_j}\hat{U}_1^{\dagger}})\hat{U}_1$ and $\langle \cdot \rangle$ is the average value under the state $\hat{S}(\xi_1)\vert\psi^{in}\rangle\vert0^{\nu}\rangle$. Substituting the explicit expression of $U_1$ and simplifying yields the QFI in the compact form:  
\begin{small} 
	\begin{equation}
			F_j(\theta_j) = \frac{4T_1^2}{1-T_1^2}\langle\hat{M}_i\rangle\times\left\vert\frac{\partial\epsilon}{\partial\theta_j}\right\vert^2,\label{eq:9}
		\end{equation}
\end{small}
where $\hat{M}_i = \hat{a}_2^\dagger \hat{a}_2$ (with $i=c,q$ denotes different input configuration), whose expectation values are taken with respect to the states $\hat{S}(\xi_1) |\psi^{\mathrm{in}}\rangle |0^\nu\rangle$. After the QFI is calculated, one can get the QCRB of the polarization angles by
\begin{small} 
	\begin{equation}
		\Delta^{QCRB}_{\theta_j}=\frac{1}{\sqrt{mF_j}},\label{eq:10}
	\end{equation}
\end{small}
where m represent the number of trials and we set $m=1$ for simplicity. Similarly, following the approach outlined in \cite{chen2025estimation}, the SQL for the polarization angles can also be derived as
\begin{small} 
	\begin{equation}
		\Delta^{SQL}_{\theta_j}=\frac{1}{\sqrt{G_j}},\label{eq:11}
	\end{equation}
\end{small}
with
\begin{small} 
	\begin{equation}
			G_j(\theta_j) = \frac{T_1^2}{1-T_1^2}\langle\hat{N}_i\rangle\times\left\vert\frac{\partial\epsilon}{\partial\theta_j}\right\vert^2,\label{eq:12}
	\end{equation}
\end{small}
where $ \hat{N}_i = \hat{a}_2^\dagger \hat{a}_2 + \hat{b}_2^\dagger \hat{b}_2$ (with $i=c,q$) is the total photon number operator entering the atomic interaction region. Its expectation values are taken with respect to the states $|\psi^{\mathrm{in}}\rangle |0^\nu\rangle$.

Following the same two input configurations discussed in Sec.~\ref{sec:subsection_ii_B}, we now evaluate the expectation values $\langle\hat{M}_i\rangle$ and $\langle\hat{N}_i\rangle$ for each case. For dual coherent state as input, where $\vert\psi^{in}\rangle=\vert\alpha\rangle\vert\beta\rangle$,
\begin{small}
	\begin{equation}
		\begin{aligned}
			\langle\hat{M}_c\rangle &=
			\bigl[|\alpha|\cosh(g)+|\beta|\sinh(g)\bigr]^2\cosh^2(g) \\
			&\quad + |\beta|^2\sinh^2(g)\bigl[2\sinh(g)+1\bigr] \\
			&\quad + |\alpha||\beta|\sinh(2g)\cosh(g) \\
			&\quad + \sinh^2(g)\bigl[\cosh(g)+1\bigr]^2, \\[4pt]
			\langle\hat{N}_c\rangle &=
			\bigl(|\alpha|^2+|\beta|^2\bigr)\cosh(2g) + 2|\alpha||\beta|\sinh(2g) \\
			&\quad + 2\sinh^2(g).
		\end{aligned}
	\end{equation}
\end{small}
For a coherent state combined with a squeezed vacuum state as input, where $\vert\psi^{in}\rangle=\vert\alpha\rangle\hat{S}(\xi)\vert0\rangle$,
\begin{small} 
	\begin{equation}
		\begin{split}
			\langle\hat{M}_q\rangle &= |\alpha|^2\cosh^4(g)+\cosh^2(r)\sinh^2(g)[\cosh(g)+1]^2,\\
			\langle\hat{N}_q\rangle &= [|\alpha|^2+\cosh^2(r)]\cosh(2g)+1\\
		\end{split}
	\end{equation}
\end{small}
By substituting these expressions back into Eqs.~\eqref{eq:9} and \eqref{eq:12}, we can concretely calculate the values of the SQL and the QCRB under the given conditions.

This is a two-parameter estimation problem ($\theta_p$ and $\theta_a$). A potential point of confusion is why we did not adopt a simultaneous estimation scheme. To clarify this point, we demonstrate the infeasibility of a simultaneous estimation scheme within our framework by analyzing the corresponding quantum Fisher information matrix (QFIM).

 We first recall that in multi-parameter quantum estimation, a sufficient condition for saturating the QCRB is the commutativity of the symmetric logarithmic derivative (SLD) operators, i.e., $[L_p, L_a] = 0$ \cite{szczykulska2016multi}. The SLD operator is implicitly defined by the following Lyapunov equation \cite{zhou2025holevo}:
\begin{small} 
	\begin{equation}
			\frac{\partial\rho_{\theta}}{\partial\theta_j}=\frac{\rho_{\theta}L_j+L_j\rho_{\theta}}{2},
	\end{equation}
\end{small}
where $\rho_{\theta}$ denotes a set of quantum states that encode all the parameters to be estimated. Specifically, for our pure state model where $\rho_{\theta}=|\psi\rangle\langle\psi|$, the Lyapunov equation admits a direct solution, yielding the SLD operator in the simple form:
\begin{small} 
	\begin{equation}
			L_j=2\frac{\partial\rho_{\theta}}{\partial\theta_j}=2(|\psi\rangle\langle\psi_j'|+|\psi_j'\rangle\langle\psi|)
	\end{equation}
\end{small}

Observing that $\frac{|\psi_p'\rangle}{\partial\epsilon/\partial\theta_p}=\frac{|\psi_a'\rangle}{\partial\epsilon/\partial\theta_a}$ and that the derivatives $\partial\epsilon/\partial\theta_p$ and $\partial\epsilon/\partial\theta_a$ commute, it follows directly that the SLD operators commute, i.e., $[L_p, L_a] = 0$. Thus, the saturation of the QCRB is guaranteed for our estimation problem and can be used safely. 

We can then define the entries of the QFIM in terms of the SLD operator as 
\begin{small} 
	\begin{equation}
	F_{jj'} = \operatorname{Tr} \left[ \rho_{\theta} \frac{L_j L_{j'} + L_{j'} L_j}{2} \right].
	\end{equation}
\end{small}
We find that the QFIM for the two-parameter simultaneous estimation is singular, thus no finite QCRB exists. This singularity arises because the measurement outcome---the single absorption index $\epsilon$---contains insufficient information to estimate both polarization angles uniquely \cite{yang2025overcoming}. In other words, a given value of $\epsilon$ corresponds not to a unique pair $(\theta_p, \theta_a)$, but to a contour line in the two-dimensional parameter space. This inherent one-to-many mapping introduces a fundamental ambiguity that precludes the simultaneous determination of both angles. Consequently, as detailed in Sec.~\ref{sec:subsection_V_A}, we circumvent this limitation by sweeping $\theta_a$ (or both $\theta_a$ and $\theta_p$), performing multiple measurements of the absorption index $\epsilon$, and thereby decomposing the problem into two sequential (or independent) single-parameter estimations. For each such single-parameter estimation, the measurement sensitivity derived in Sec.~\ref{sec:subsection_ii_B}, along with the corresponding QCRB and SQL given by Eqs.~(\ref{eq:10}) and (\ref{eq:11}), can be applied.

\section{PRINCIPLE OF RYDBERG ATOM-BASED POLARIMETRY}\label{sec:section_iV}
\subsection{Description of the polarization }\label{sec:subsection_iV_A}
\begin{figure*}[tbp]
	\label{Fig2} \centering \includegraphics[width=1\textwidth]{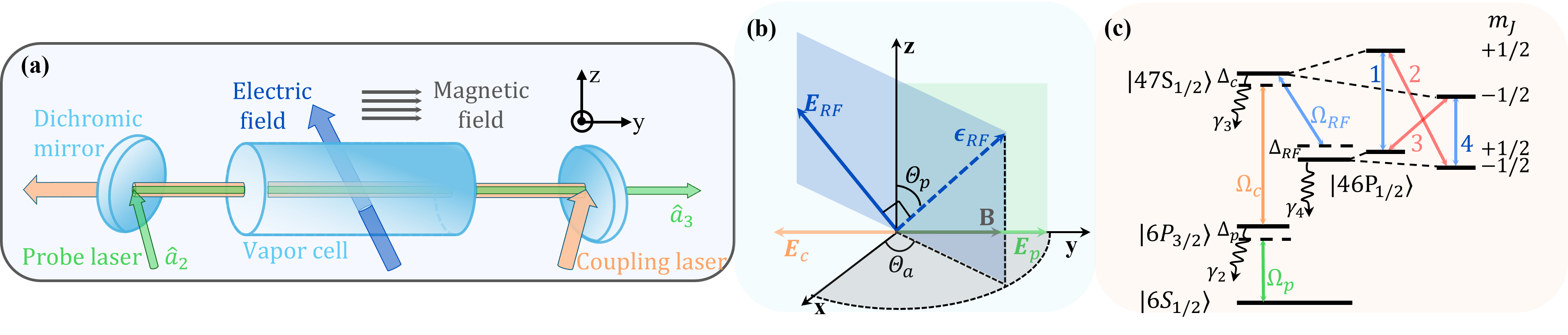}%
	\newline
	\caption{(a) Schematic diagram of the RAP. The probe mode, corresponding to input mode $\hat{a}_2$ of BS1, propagates through the vapor cell and exits as mode $\hat{a}_3$ after interacting with the Rydberg atoms. The coupling laser counter-propagates relative to the probe mode. A static magnetic field is applied along the $y$-axis. A vector electric field is incident from an arbitrary direction and couples to the Rydberg atoms within the vapor cell. (b) Representation of the atomic system in the quantization axis specified by the static magnetic field $\textbf{B}$. The polarization vector of the RF electric field $\boldsymbol{\epsilon}_{\mathrm{RF}}$ is defined by the polar angle $\theta_p$ and the azimuthal angle $\theta_a$. (c) Schematic of the $\Lambda$-type four-level Rydberg system under magnetic field control. The Zeeman sublevels of the two Rydberg states give rise to four possible transition pathways, as indicated in the diagram. Pathways coupled by the linear-polarized components of the field are shown in blue, and those coupled by the circular-polarized components are shown in red.}
	\label{fig:2}
\end{figure*}
A schematic of the RAP setup is shown in Figs.~\ref{fig:2}(a) and (b), with the quantization axis defined by a static magnetic field $\mathbf{B}$ applied along the y-axis. The probe mode and coupling laser beams counter-propagate along the $y$-axis, while the electric field $\mathbf{E}_{\mathrm{RF}}$ is incident from an arbitrary direction. The weak static magnetic field $\mathbf{B}$ lifts the degeneracy of the Rydberg levels. The polarization vector $\boldsymbol{\epsilon}_{\mathrm{RF}}$ is initially defined in Cartesian coordinates by a polar angle $\theta_p$ and an azimuthal angle $\theta_a$ [see Fig.~\ref{fig:2}(b)]. However, to analyze its interaction with the atomic dipole moments, it is more convenient to decompose it in the standard spherical basis $\{\boldsymbol{\epsilon}_{-1}, \boldsymbol{\epsilon}_0, \boldsymbol{\epsilon}_{+1}\}$, where $\boldsymbol{\epsilon}_0=\hat{y}$ and $\boldsymbol{\epsilon}_{\pm1}=({\mp\hat{x}-i\hat{z}})/{\sqrt{2}}$. In this basis, $\boldsymbol{\epsilon}_0$ represents the $\pi$-polarized unit vector driving $\Delta m = 0$ transitions, while $\boldsymbol{\epsilon}_{\pm 1}$ correspond to the $\sigma^{\pm}$-circularly polarized unit vectors driving $\Delta m = \pm 1$ transitions. Consequently, the electric field can be expressed as
\begin{small}
	\begin{equation}
		\begin{split}
			\mathbf{E}_{\text{RF}} &= E_{\text{RF}} \left( \cos\theta_a \sin\theta_p \, \hat{\mathbf{x}} + \sin\theta_a \sin\theta_p \, \hat{\mathbf{y}} + \cos\theta_p \, \hat{\mathbf{z}} \right) \\
			& = E_{\text{RF}}\sum_{q=-1}^{+1}\alpha_q \boldsymbol{\epsilon}_q =\sum_{q=-1}^{+1}\mathbf{E}^{(q)}_{\text{RF}},
		\end{split}\label{eq:18}
	\end{equation}
\end{small}
with
\begin{small} 
	\begin{equation}
		\begin{split}
			\alpha_0 &= \sin\theta_a \sin\theta_p, \\
			\alpha_{\pm 1} &= \mp\frac{1}{\sqrt{2}} \sqrt{\cos^2\theta_p + \sin^2\theta_p \cos^2\theta_a} \, e^{\mp i \theta'}, \\
			\theta' &= \arctan\left( {\cot\theta_p\sec\theta_a } \right),
		\end{split}\label{eq:19}
	\end{equation}
\end{small}

where $E_{\mathrm{RF}}$ is the electric field amplitude, and $\alpha_q$ ($q=0, \pm 1$) are the dimensionless projection coefficients of the polarization vector $\boldsymbol{\epsilon}_{\mathrm{RF}}$ onto the spherical basis vectors $\boldsymbol{\epsilon}_q$. Their squared moduli, $|\alpha_q|^2$, represent the relative intensity of each polarization component ($\pi$ or $\sigma^{\pm}$). These coefficients depend solely on the polarization angles $\theta_a$ and $\theta_p$. Notably, the phase angle $\theta'$ does not affect the total susceptibility $\chi$; it cancels out in the final expression because the susceptibility, within the first-order perturbation approximation, depends only on the products $\alpha_q^* \alpha_q = |\alpha_q|^2$, as will be shown in the next subsection.

\subsection{Interaction with a four-level Rydberg atom system}\label{sec:subsection_iV_B}
The $\Lambda$-type four-level Rydberg system under magnetic-field control is depicted in Fig.~\ref{fig:2}(c). Compared to a ladder-type configuration, this $\Lambda$ scheme offers a larger transition dipole moment, making it particularly suitable for weak-field sensing. We select the Rydberg transition $|n'S_{1/2}\rangle \to |nP_{1/2}\rangle$ because its energy eigenvalues are invariant under rotations of the RF field polarization \cite{cloutman2024polarization}. Consequently, the Autler–Townes (AT) splitting interval remains unchanged as the polarization varies. This crucial feature enables the independent measurement of the field polarization and amplitude without mutual interference.

The RF electric field couples the two Rydberg states, whose energies are tuned by a weak static magnetic field $\mathbf{B}$. For simplicity, we adopt two key approximations when incorporating the magnetic field: (i) The magnetic-field-induced shifts of the ground state $|6S_{1/2}\rangle$ and the intermediate state $|6P_{3/2}\rangle$ are neglected. This is justified because these shifts are small and can be compensated by tuning the frequencies of the probe mode and coupling laser. (ii) The hyperfine structure of the Rydberg states is omitted. This is valid since, for the applied magnetic field strength ($\sim 1\ \mathrm{G}$), the hyperfine splitting ($\sim \mathrm{kHz}$) is negligible compared to the Zeeman shift ($\sim \mathrm{MHz}$).

From this simplified model, the possible transitions between the Rydberg states arise from pairwise combinations within the sets $\{|n'S_{1/2}, m_J' = +1/2\rangle, |n'S_{1/2}, m_J' = -1/2\rangle\}$ and $\{|nP_{1/2}, m_J = +1/2\rangle, |nP_{1/2}, m_J = -1/2\rangle\}$, resulting in four distinct transition pathways, labeled by an index $p$ ($p = 1, 2, 3, 4$), as is shown in Fig.~\ref{fig:2}(c). Transitions with $\Delta m = 0$ ($\pi$-transitions) couple to the $\boldsymbol{\epsilon}_0$ (linear) component of the electric field, while those with $\Delta m = \pm 1$ ($\sigma^{\pm}$-transitions) couple to the $\boldsymbol{\epsilon}_{\pm 1}$ (circular) components, respectively.

Furthermore, the detuning for each transition pathway is magnetically tunable and can be expressed as
\begin{small}
	\begin{equation}
		\begin{split}
			\Delta^{(q)'}_{\mathrm{RF}} &= \Delta_{\mathrm{RF}} + \Delta^{(q)}_B, \\
			\Delta^{(q)}_B &= \mu_B B \left( g_{J, r_1} m_J'^{\,(q)}-g_{J, r_2} m_J^{(q)}   \right),
		\end{split}
	\end{equation}
\end{small}
where $\Delta_{\mathrm{RF}}$ is the detuning in the absence of the magnetic field, $\Delta_B^{(q)}$ is the magnetically induced shift, $\mu_B$ is the Bohr magneton, $B$ is the magnetic field strength, and $g_{J, r_1}$ and $g_{J, r_2}$ are the Land\'e $g$-factors for the $|n'S_{1/2}\rangle$ and $|nP_{1/2}\rangle$ states, respectively. The magnetic quantum numbers $m_J'^{\,(q)}$ and $m_J^{(q)}$ correspond to the specific Zeeman sublevels involved in the $q$-th polarization component.

Within the electric-dipole approximation, the atom-field interaction is described by expanding the dipole operator $\hat{\mathbf{d}}$ in the spherical tensor basis of rank 1:
\begin{small}
	\begin{equation}
		\hat{\mathbf{d}} = \sum_{q=-1}^{1} \hat{d}_q \, \boldsymbol{\epsilon}_q.
	\end{equation}
\end{small}
For a transition pathway characterized by a spherical component index $q$ (i.e., $\Delta m = q$), the Rabi frequency is defined as
\begin{small}
	\begin{equation}
		\begin{split}
			\Omega^{(q)}_{\mathrm{RF}}(\theta_p, \theta_a) &= -\frac{1}{\hbar} \langle nP_{1/2}, m_J | \, \hat{\mathbf{d}} \cdot \mathbf{E}^{(q)}_{\mathrm{RF}} \, | n'S_{1/2}, m_J' \rangle \\
			&= -\frac{\alpha_q E_{\mathrm{RF}}}{\hbar} \langle nP_{1/2}, m_J | \hat{d}_q | n'S_{1/2}, m_J' \rangle,
		\end{split}
	\end{equation}
\end{small}
where $\mathbf{E}^{(q)}_{\mathrm{RF}} = \alpha_q E_{\mathrm{RF}} \boldsymbol{\epsilon}_q$ is the component of the electric field along the spherical basis vector $\boldsymbol{\epsilon}_q$. The simplification from the first to the second line follows from the orthonormality of the spherical basis: $\boldsymbol{\epsilon}_q^* \cdot \boldsymbol{\epsilon}_{q'} = \delta_{qq'}$. A key consequence is that all transition pathways associated with the same spherical component $q$ (and thus the same $\Delta m$) exhibit identical Rabi frequencies, $\Omega^{(q)}_{\mathrm{RF}}$. Applying the Wigner–Eckart theorem, the matrix element can be decomposed into a product of a Clebsch--Gordan coefficient and a direction-independent reduced matrix element\cite{steck2007quantum}.
\begin{small}
	\begin{equation}
		\begin{split}
		&\langle nP_{1/2}, m_J | \hat{d_q} | n'S_{1/2}, m_J' \rangle \\
		&= \langle \tfrac{1}{2} \, m_J' | \tfrac{1}{2} \, m_J; 1 \, q \rangle \, \langle nP_{1/2} \| \hat{\mathbf{d}} \| n'S_{1/2} \rangle\\
		&=(-1)^{\tfrac{1}{2} - m_J} \times\sqrt{2}D \,
		\begin{pmatrix}
			1/2 & 1 & 1/2 \\
			m_J & q & -m_J'
		\end{pmatrix},
		\end{split}
	\end{equation}
\end{small}
where the third line expresses the Clebsch–Gordan coefficient in terms of a Wigner 3j-symbol. Here, $D=\langle nP_{1/2} \| \hat{\mathbf{d}} \| n'S_{1/2} \rangle$ is the radial matrix element for the Rydberg transition under consideration. Thus, the final expression for the Rabi frequency of the transition pathway is given by:
\begin{small}
	\begin{equation}
		\Omega^{(q)}_{\mathrm{RF}}(\theta_p,\theta_a) = \frac{(-1)^{\tfrac{3}{2} - m_J}\sqrt{2} \, \alpha_q \, D \, E_{\mathrm{RF}}}{\hbar}  \,
		\begin{pmatrix}
			1/2 & 1 & 1/2 \\
			m_J & q & -m_J'
		\end{pmatrix},\label{eq:24}
	\end{equation}
\end{small}
Within the dipole and rotating-wave approximations, the Hamiltonian in the interaction picture for a given transition pathway p(associated with a spherical component q) is
\begin{small}
	\begin{equation}
		\hat{H}_I^{(p)} = \hbar
		\begin{pmatrix}
			0 & \xi^* \hat{a}_2^\dagger & 0 & 0 \\[2pt]
			\xi \hat{a}_2 & \Delta_p & \frac{\Omega_c^*}{2} & 0 \\[2pt]
			0 & \frac{\Omega_c}{2} & \Delta_p + \Delta_c & \frac{\Omega_{\mathrm{RF}}^{*\,(q)}}{2} \\[2pt]
			0 & 0 & \frac{\Omega_{\mathrm{RF}}^{(q)}}{2} & \Delta_p + \Delta_c + \Delta_{\mathrm{RF}}^{(p)'}
		\end{pmatrix},
	\end{equation}
\end{small}
where the operator $\hat{a}_2$ represents one of the input mode of BS1 in the SU(1,1) interferometer. The coupling constant between the probe field and the atomic transition $|1\rangle \leftrightarrow |2\rangle$ is $\xi = \mu_{12} \varepsilon / \hbar$, with $\mu_{ij}$ denoting the transition dipole moment between states $|i\rangle$ and $|j\rangle$, $\varepsilon = \sqrt{\hbar \omega_p / (2 \epsilon_0 V)}$ the single-photon electric field amplitude, $V$ is the quantization volume, and $\epsilon_0$ is the vacuum permittivity. The coupling laser and the RF electric field are treated classically, with their corresponding Rabi frequencies given by $\Omega_c = \mu_{23} E_c / \hbar$ and $\Omega^{(q)}_{\mathrm{RF}}$, respectively. Here, $\hbar$ is the reduced Planck constant, and $\Delta_p$ and $\Delta_c$ denote the detunings of the probe mode and coupling laser. The spherical component index $q$ is determined by the magnetic quantum number change $\Delta m$ associated with the transition pathway under consideration.

To account for decoherence processes such as spontaneous emission, collisions, and transit-time broadening, the dynamics of the atomic operators for the $p$-th transition pathway are governed by the Heisenberg--Langevin equation \cite{PhysRevApplied.20.064028}: 

\begin{small}
	\begin{equation}
		\frac{\partial}{\partial t} \hat{\sigma}_{\mu\nu}^{(p)} = \frac{1}{i\hbar} \left[ \hat{H}_I^{(p)}, \, \hat{\sigma}_{\mu\nu}^{(p)} \right] + \hat{D}_{\mu\nu} + \hat{F}_{\mu\nu},
	\end{equation}
\end{small}
where $\hat{\sigma}_{\mu\nu}^{(p)}(z,t) = \frac{1}{N_z} \sum_{j=1}^{N_z} |\mu_j\rangle\langle\nu_j| \, e^{-i\Delta_p t + i k_p z}$ is the collective slowly varying operator that represents an average over $N_z$ atoms within a thin layer at position $z$. $k_p$ is the wave vector of the probe light, and $\hat{F}_{\mu\nu}$ is the corresponding Langevin noise operator and $\hat{D}_{\mu\nu}$ represents the dissipator matrix which can be derived by
	
\begin{small}
	\begin{equation}
		\hat{D}_{\mu\nu} =
		\begin{pmatrix}
			\gamma_2 \hat{\sigma}_{22} + \gamma_4 \hat{\sigma}_{44} & -\frac{\gamma_2}{2}\hat{\sigma}_{12} & -\frac{\gamma_3}{2}\hat{\sigma}_{13} & -\frac{\gamma_4}{2}\hat{\sigma}_{14} \\[4pt]
			-\frac{\gamma_2}{2}\hat{\sigma}_{21} &  \gamma_3 \hat{\sigma}_{33} -\gamma_2 \hat{\sigma}_{22} & -\frac{\gamma_{23}}{2}\hat{\sigma}_{23} & -\frac{\gamma_{24}}{2}\hat{\sigma}_{24} \\[4pt]
			-\frac{\gamma_3}{2}\hat{\sigma}_{31} & -\frac{\gamma_{32}}{2}\hat{\sigma}_{32} & -\gamma_3 \hat{\sigma}_{33} & -\frac{\gamma_{34}}{2}\hat{\sigma}_{34} \\[4pt]
			-\frac{\gamma_4}{2}\hat{\sigma}_{41} & -\frac{\gamma_{42}}{2}\hat{\sigma}_{42} & -\frac{\gamma_{43}}{2}\hat{\sigma}_{43} & -\gamma_4 \hat{\sigma}_{44}
		\end{pmatrix},
	\end{equation}
\end{small}
where $\gamma_{mn} \equiv \gamma_m + \gamma_n$, and $\gamma_m$ ($m=2,3,4$) is the spontaneous decay rate (see Fig. \ref{fig:2}c).

The propagation and temporal evolution of the probe field operator $\hat{a}_2$ are governed by
\begin{small}
	\begin{equation}
		\left( \frac{\partial}{\partial t} + c \frac{\partial}{\partial z} \right) \hat{a}_2(z,t) = i \xi \mathcal{N} \hat{\sigma}_{12}^{(p)}(z,t),
	\end{equation}
\end{small}
where $\mathcal{N}$ is the atomic density and $c$ is the speed of light in vacuum. Taking the Fourier transform with respect to time and space ($\partial/\partial t \rightarrow i\omega$, $\partial/\partial z \rightarrow i k_p$) yields
\begin{small}
	\begin{equation}
		i\omega \hat{a}_2(\omega, k) + i c k_p \hat{a}_2(\omega, k) = i \xi \mathcal{N} \hat{\sigma}_{12}^{(p)}(\omega, k).
	\end{equation}
\end{small}
Rearranging terms and using the relation $c k_p = \omega n^{(p)}$ with the refractive index $n^{(p)} = \sqrt{1 + \chi^{(p)}}$, and taking the average of the operators on both sides, we obtain the susceptibility of the Rydberg atomic medium associated with the $p$-th transition pathway as
\begin{small}
	\begin{equation}
		\begin{split}
		&c k_p \left( 1 + \frac{1}{\sqrt{1+\chi^{(p)}}} \right) \langle\hat{a}_2\rangle = \xi \mathcal{N} \langle\hat{\sigma}_{12}^{(p)}\rangle \quad \\
		&\Longrightarrow \quad \chi^{(p)} = \frac{1}{\left(1 - \dfrac{\xi \mathcal{N}}{c k_p} \dfrac{\langle\hat{\sigma}_{12}^{(p)}\rangle}{\langle\hat{a}_2\rangle}\right)^2 } - 1 \; \approx \; \frac{2 \xi \mathcal{N}}{c k_p} \frac{\langle\hat{\sigma}_{12}^{(p)}\rangle}{\langle\hat{a}_2\rangle},
	\end{split}\label{eq:30}
	\end{equation}
\end{small}
where the last approximation utilizes the fact that $(\xi \mathcal{N}\langle\hat{\sigma}_{12}^{(p)}\rangle )/ (c k_p\langle\hat{a}_2\rangle) \ll 1$. Solving the Heisenberg-Langevin equations within the first-order perturbation approximation for $\langle \hat{\sigma}_{12}^{(p)} \rangle$ and substituting the result into Eq.~(\ref{eq:30}) yields the explicit susceptibility for the $p$-th transition pathway:
\begin{small}
	\begin{equation}
			\chi^{(p)}=\frac{i\mathcal{N}\left|\mu_{21}\right|^{2}}{\hbar\epsilon_{0}}\frac{\Gamma_{13}\Gamma_{14}^{(p)}+\frac{\Omega_{\mathrm{RF}}^{(q)}\Omega_{\mathrm{RF}}^{*(q)}}{4}}{\Gamma_{12}\left[\Gamma_{13}\Gamma_{14}^{(p)}+\frac{\Omega_{\mathrm{RF}}^{(q)}\Omega_{\mathrm{RF}}^{*(q)}}{4}\right]+\frac{\Omega_{c}\Omega_{c}^{*}}{4}},\label{eq:31}
	\end{equation}
\end{small}
where $\Gamma_{12}=i\Delta_p + {\gamma_2}/{2} $, $\Gamma_{13}=i(\Delta_p + \Delta_c) + {\gamma_3}/{2}$ and $\Gamma_{14}^{(p)} = i(\Delta_p + \Delta_c + \Delta_{\mathrm{RF}}^{(p)'}) +{\gamma_4}/{2}$. Notably, $\chi^{(p)}$ depends on the RF field only through the product $\Omega_{\mathrm{RF}}^{(q)}\Omega_{\mathrm{RF}}^{*(q)} = |\Omega_{\mathrm{RF}}^{(q)}|^2$. Because $\Omega_{\mathrm{RF}}^{(q)} \propto \alpha_q$ [see Eq.~\eqref{eq:24}], this implies that $\chi^{(p)}$ depends solely on $|\alpha_q|^2 = \alpha_q^*\alpha_q$. Consequently, the phase angle $\theta'$ introduced in Eq.~\eqref{eq:19} does not affect the susceptibility, consistent with the earlier discussion in Sec.~\ref{sec:subsection_iV_A}. The total susceptibility $\chi$ of the medium is the sum over all four transition pathways. The corresponding absorption index $\epsilon$ and phase shift $\theta$ of the probe field after traversing the medium are given by

\begin{small}
	\begin{equation}
		\begin{split}
		\chi &= \sum_{p=1}^{4} \chi^{(p)}(\theta_p, \theta_a), \\
		\epsilon &= \frac{2\pi l}{\lambda_p} \operatorname{Im}[\chi],~  
		\theta = \frac{\pi l}{\lambda_p} \operatorname{Re}[\chi],
		\end{split}\label{eq:32}
	\end{equation}
\end{small}
where $l$ is the length of the atomic vapor cell and $\lambda_p$ is the wavelength of the probe mode.

Thus, we have derived a theoretical model that quantitatively describes the response of the Rydberg atomic medium to the probe beam under the joint modulation of a linearly polarized RF field and a coupling laser. 
Consequently, by utilizing the SU(1,1) interferometer to perform precise measurements of the absorption index $\epsilon$, the polarization angles $\theta_p$ and $\theta_a$ of the incident RF field can be extracted.

\section{RESULT}\label{sec:section_V}

\subsection{Absorption index versus polarization angles }\label{sec:subsection_V_A}
\begin{figure}[tbp]
	\label{Fig3} \centering \includegraphics[width=1\columnwidth]{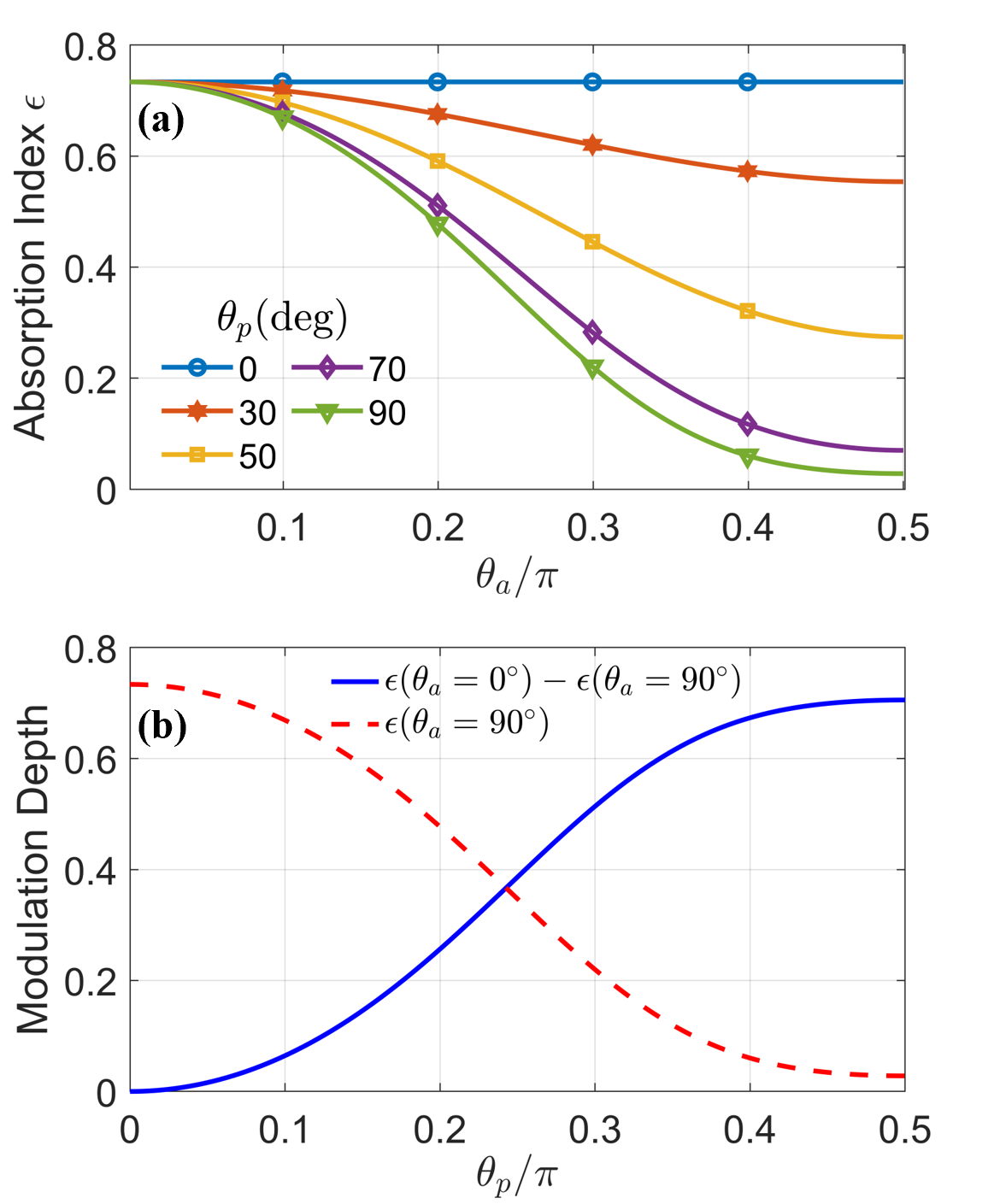}%
	\newline
	\caption{(a) Absorption index $\epsilon$ as a function of polarization angles $\theta_a$ and $\theta_p$. Each curve corresponds to a fixed value of $\theta_p$, plotting $\epsilon$ versus $\theta_a$. (b) The variation amplitude of $\epsilon$ over one period of $\pi$, defined as the modulation depth, plotted as a function of $\theta_p$ (blue solid line). Also shown is the absorption index at $\theta_a = 90^\circ$ as a function of $\theta_p$ (red dashed line).}
	\label{fig:3}
\end{figure}
For the numerical simulation, we consider Cs atoms in a vapor cell which contains ground-state atoms at a total density of $\mathcal{N}=4.9\times10^{16}\,\mathrm{m}^{-3}$. The $852\,\text{nm}$ probe mode excites the ground state $|6S_{1/2}\rangle$ to the intermediate state $|6P_{3/2}\rangle$, which is subsequently excited to the Rydberg state $|47S_{1/2}\rangle$ by a $510\,\text{nm}$ coupling laser. The transition between this state and a neighboring Rydberg state $|46P_{1/2}\rangle$ is coupled to the RF electric field. For the $|6S_{1/2}\rangle \to |6P_{3/2}\rangle$ transition, the decay rate and transition dipole element are $\gamma_2 = 32.89 \times 10^6\,\text{s}^{-1}$ and $\mu_{12} = 4.48\, e a_0$, respectively \cite{steck2003cesium}. For the two Rydberg states, the decay rates are $\gamma_3 = 3.23 \times 10^3\,\text{s}^{-1}$ and $\gamma_4 = 0.14 \times 10^3\,\text{s}^{-1}$, and the radial matrix element for the Rydberg transition  $D = 1880.39\, e a_0$, which are computed using the Alkali-Rydberg-Calculator (ARC) package \cite{robertson2021arc}, a widely-used and experimentally benchmarked library for calculating properties of alkali and divalent atoms. Pioneering work of Ref.~\cite{sedlacek2013atom} demonstrated polarization measurement of microwave fields with amplitudes  $>1\,\mathrm{V/m}$. In contrast, to highlight that our scheme is suitable for weak-field detection, we set the RF electric field amplitude to $0.1\,\mathrm{V/m}$. Under this condition, to ensure that the absorption index varies monotonically with $\theta_j$ over the range $[0, \pi/2]$ and to achieve optimal sensitivity, we select the static magnetic field strength to $B = 0.56\,\mathrm{Gs}$ and the coupling detuning to $\Delta_c = -2\pi\,\mathrm{MHz}$. The values of these two parameters are determined through subsequent parameter optimization, as detailed in subsection \ref{sec:subsection_V_C} 

Figure~\ref{fig:3}(a) shows the absorption index $\epsilon$ as a function of $\theta_a$ and $\theta_p$ under the chosen parameters. For any fixed $\theta_p$, $\epsilon$ varies monotonically with $\theta_a$ over the interval $[0, \pi/2]$. Moreover, $\epsilon$ is periodic in $\theta_a$ with period $\pi$, and its behavior on $[\pi/2, \pi]$ mirrors that on $[0, \pi/2]$ with respect to $\theta_a = \pi/2$. In Fig.~\ref{fig:3}(b), the modulation depth (MD), defined as the variation amplitude of $\epsilon$ over one period $\pi$, is plotted as a blue solid line. The MD depends only on $\theta_p$ and varies monotonically with $\theta_p$ over the range $[0, \pi/2]$, which can be expressed as $MD = \epsilon(\theta_a=0^\circ) - \epsilon(\theta_a=90^\circ)$.

A key observation from Eqs.~\eqref{eq:18}, \eqref{eq:19}, \eqref{eq:24}, \eqref{eq:31}, and \eqref{eq:32} is that $\epsilon(\theta_a=0^\circ)$ is independent of $\theta_p$ (a feature also visible in Fig.~\ref{fig:3}(a)). This allows us to introduce a refined quantity, $MD' \equiv \epsilon(\theta_a=0^\circ) - MD = \epsilon(\theta_a=90^\circ)$, which is plotted as a red dashed line in Fig.~\ref{fig:3}(b). Since $MD'$ equals $\epsilon(\theta_a=90^\circ)$, it can be obtained directly from a single absorption measurement at $\theta_a = 90^\circ$, thereby reducing the uncertainty associated with extracting the full MD.

\begin{figure}[tbp]
	\centering
	\includegraphics[width=1\columnwidth]{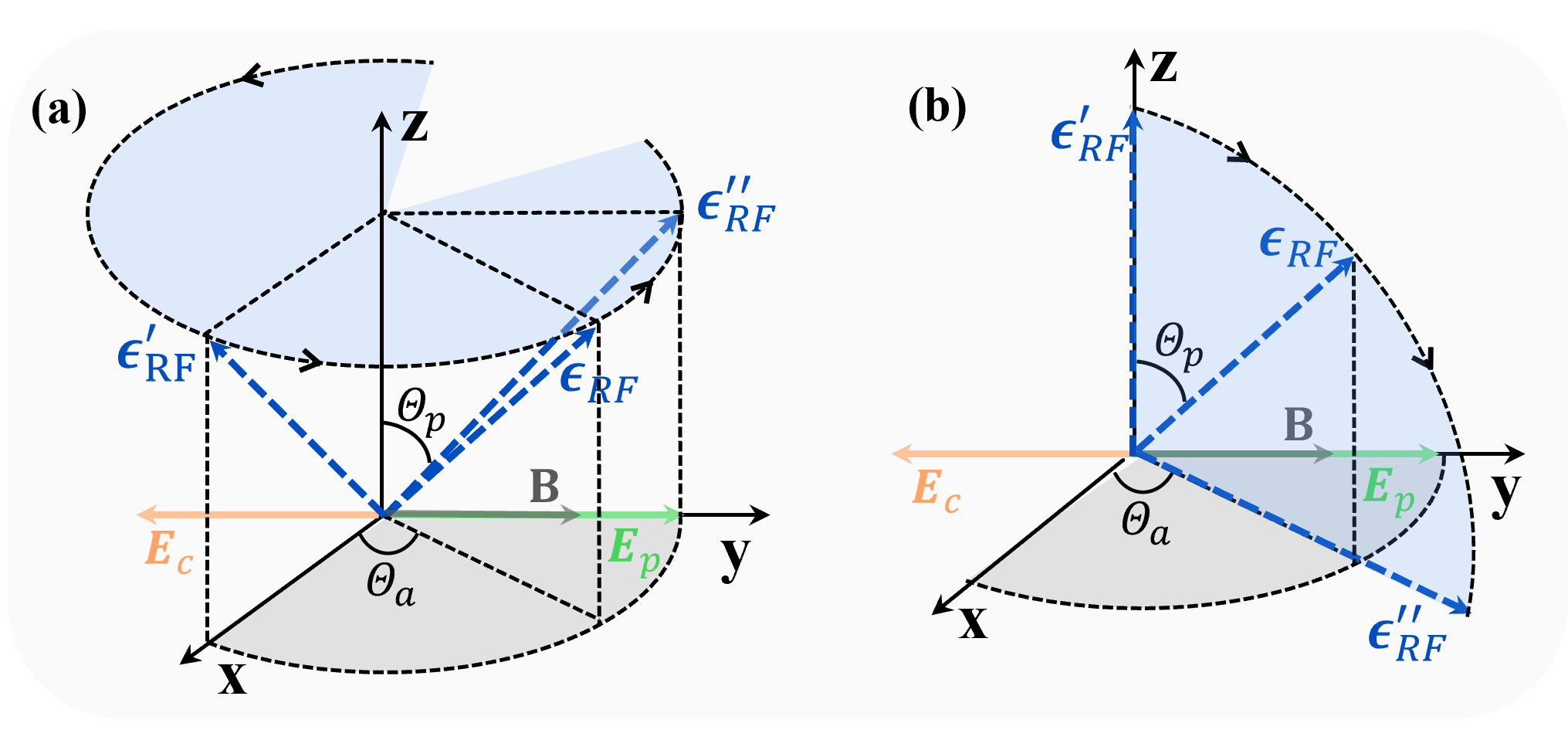}
	\caption{(a) Rotation of the polarization vector $\boldsymbol{\epsilon}_{\mathrm{RF}}$ about the $z$-axis, varying $\theta_a$ while $\theta_p$ is held constant. The initial orientation is labeled $\boldsymbol{\epsilon}_{\mathrm{RF}}$. The special orientations $\boldsymbol{\epsilon}'_{\mathrm{RF}}$ and $\boldsymbol{\epsilon}''_{\mathrm{RF}}$ correspond to $\theta_a = 0^\circ$ and $\theta_a = 90^\circ$, respectively, yielding the absorption indices $\epsilon(\theta_a=0^\circ)$ and $\epsilon(\theta_a=90^\circ)$. (b) Rotation of $\boldsymbol{\epsilon}_{\mathrm{RF}}$ within the plane spanned by the $z$-axis and $\boldsymbol{\epsilon}_{\mathrm{RF}}$ itself, varying $\theta_p$ while $\theta_a$ is fixed. Here, $\boldsymbol{\epsilon}'_{\mathrm{RF}}$ and $\boldsymbol{\epsilon}''_{\mathrm{RF}}$ denote the orientations for $\theta_p = 0^\circ$ and $\theta_p = 90^\circ$, respectively.}
	\label{fig:4}
\end{figure}

As established in Sec.~\ref{sec:section_iii}, a single measurement of $\epsilon$ does not provide sufficient information to uniquely determine both polarization angles, rendering simultaneous two-parameter estimation infeasible. To overcome this limitation, we propose two practical measurement strategies.

(i) Sequential estimation with error propagation. Initially, the polarization vector is oriented as shown by $\boldsymbol{\epsilon}_{\mathrm{RF}}$ in Fig.~\ref{fig:4}(a). A first homodyne detection yields the value $\epsilon_I$. The polarization vector $\boldsymbol{\epsilon}_{\mathrm{RF}}$ is then rotated around the z-axis to sweep $\theta_a$ while keeping $\theta_p$ fixed. This rotation can be implemented either by physically rotating the horn antenna that generates the RF field or by rotating the entire Rydberg sensor system. When the polarization vector rotates to $\boldsymbol{\epsilon}''_{\mathrm{RF}}$, $\theta_a = 90^\circ$ and the absorption index reaches its minimum, denoted as $\epsilon_M$, which is also the modulation depth $MD$. This value $\epsilon_M$ is then used to estimate $\theta_p$. The estimation error $\Delta\theta_p$ for this step is given by Eq.~\eqref{eq:3}, and the corresponding QCRB and SQL are given by Eqs.~\eqref{eq:10} and \eqref{eq:11}, respectively. Subsequently, the estimated value $\theta^{(0)}_p$ (from $\epsilon_M$) and the initially measured $\epsilon_I$ are used to estimate $\theta_a$. The error $\Delta'\theta_a$ in this second step has two sources: the measurement error of $\epsilon_I$ and the propagation of the estimation error from $\theta^{(0)}_p$. The total error in estimating $\theta_a$ is therefore $\Delta'\theta_a = \sqrt{ \Delta^2\theta_a + (\partial\epsilon/\partial\theta_p)^2 \Delta^2\theta_p/ (\partial\epsilon/\partial\theta_a)^2 } = \sqrt{2}\,\Delta\theta_a$, where $\Delta\theta_a$ is again given by Eq.~\eqref{eq:3}. The corresponding QCRB and SQL for this step are likewise $\sqrt(2)$ times those given by Eqs.~\eqref{eq:10} and \eqref{eq:11}.

(ii) Independent estimation via a second rotation. The first step for estimating $\theta_p$ is identical to Method (i). The difference lies in the estimation of $\theta_a$. Here, we employ the same technique used for estimating $\theta_p$: we rotate the polarization vector within the plane spanned by the z-axis and $\boldsymbol{\epsilon}_{\mathrm{RF}}$ itself, as illustrated in Fig.~\ref{fig:4}(b). This rotation varies $\theta_p$ while keeping $\theta_a$ fixed. Similarly, the MD extracted from this rotation is used to estimate $\theta_a$. In this method, the estimation error for $\theta_a$ is not affected by the error in the previously estimated $\theta_p$. Consequently, the error is simply $\Delta\theta_a$.

In summary, Method (i) requires a single rotation of the polarization vector and involves two sequential single-parameter estimations. In this approach, the estimation error of $\theta_p$ propagates to $\theta_a$, increasing the final error in $\theta_a$ by a factor of $\sqrt{2}$. In contrast, Method (ii) requires two separate rotations, enabling two independent single-parameter estimations. in this case, the estimates of $\theta_p$ and $\theta_a$ are decoupled, and their errors do not affect one another. Both methods require at least two measurements to determine the two angles. In the subsequent sensitivity analysis, we adopt Method (ii) as the default because the sensitivity for Method (i) can be obtained simply by scaling the $\theta_a$ sensitivity of Method (ii) by a factor of $\sqrt{2}$.
 
Furthermore, we find that our scheme is well-suited for weak-field detection. This is because our method relies solely on measuring the absorption index of the Rydberg atomic medium to determine the polarization angles. Consequently, the RF field does not need to be strong enough to fully resolve the Autler-Townes (AT) splitting peaks. This key feature permits operation at substantially lower field strengths.
 
\subsection{Sensitivity for two input configurations}\label{sec:subsection_V_B}
 We first analyze the configuration with dual coherent-state inputs, where the modes $\hat{a}_{\mathrm{in}}$ and $\hat{b}_{\mathrm{in}}$ are prepared in coherent states $|\alpha\rangle$ and $|\beta\rangle$, respectively, with $\alpha = |\alpha|e^{i\theta_\alpha}$ and $\beta = |\beta|e^{i\theta_\beta}$. The parameters for the OPA are $g = 0.7$, $\theta_{\alpha}=0$, and $\theta_{\beta}=\pi$. The coherent amplitude $|\alpha|=|\beta|=5\times10^6$ are chosen to optimize the sensitivity of the RAP by balancing the collision rates and the power broadening associated with increased laser power \cite{PhysRevApplied.20.064028}.
 
 \begin{figure}[tbp]
 	\label{Fig5} \centering \includegraphics[width=1\columnwidth]{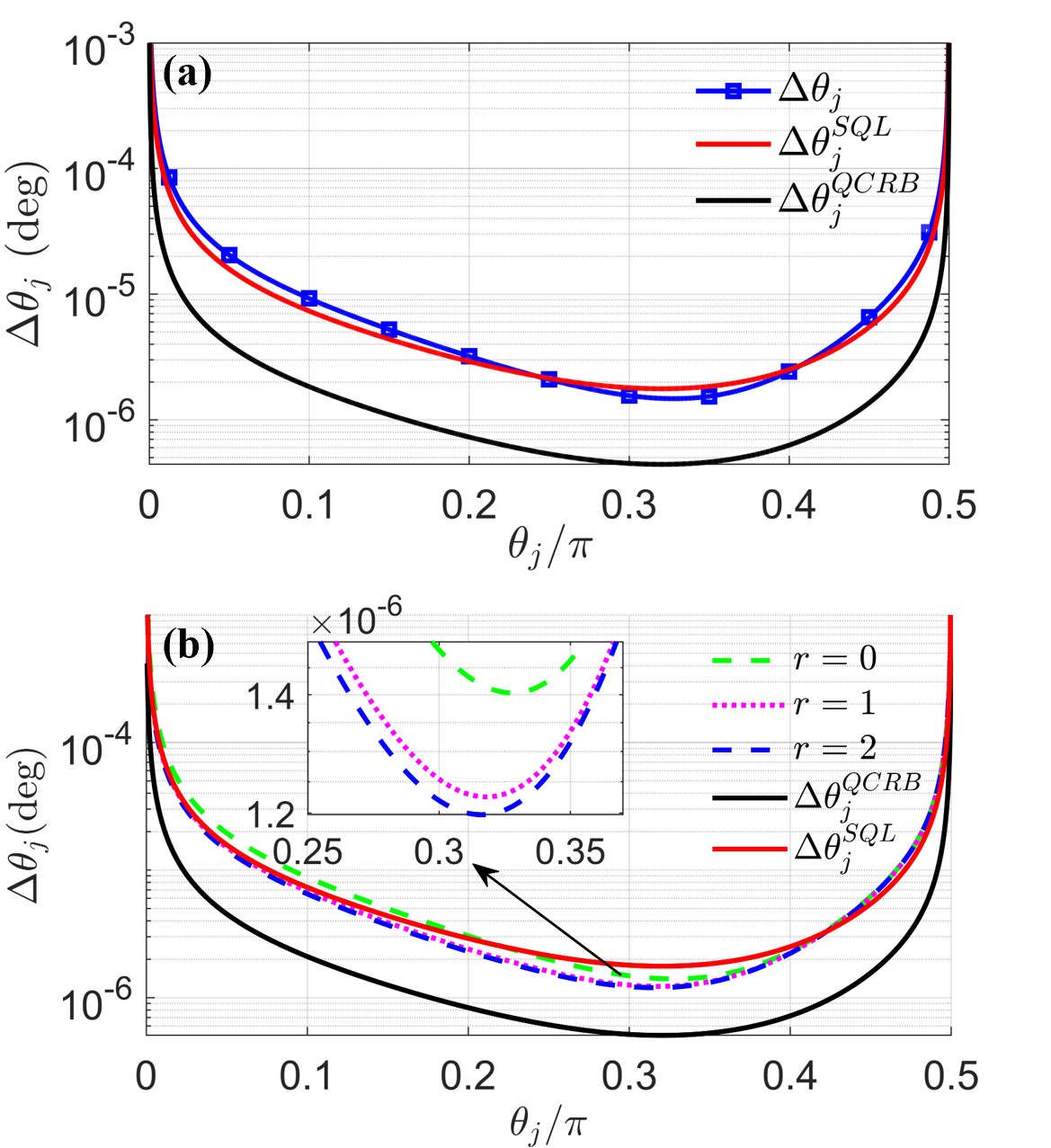}%
 	\newline
 	\caption{(a) With \textit{dual} coherent states as input, the SQL (red solid line) and QCRB (black solid line) of the polarization angle sensitivity and the sensitivity based on homodyne detection (blue solid line with square markers) as a function of $\theta_j(j=a,p)$  (b) With a coherent state combined with a squeezed vacuum state as input, the SQL (red solid line) and QCRB (black solid line) of the Polarization angle sensitivity and the sensitivity based on homodyne detection for squeezed parameter $r=0$ (green dash line), $r=1$ (magenta dotted line) and $r=2$ (blue dash line) as a function of $\theta_j$.  
 	}	\label{fig:5}
 \end{figure}
 
  Fig.~\ref{fig:5}(a) shows the polarization angle sensitivity achieved via homodyne detection (blue solid line with squares) as a function of $\theta_j$ $(j=a,p)$. TFor comparison, the corresponding SQL (red solid line) and QCRB (black solid line) are also plotted, illustrating the quantum enhancement provided by our scheme. It is clearly seen that the sensitivity $\Delta\theta_j$ remains below \SI{e-3}{\degree} over most of the angular domain, excluding regions near $0$ and $\pi/2$, where the slope $|\partial\epsilon/\partial\theta_j|$ approaches zero. The optimal sensitivity of \SI{1.47e-6}{\degree} is attained around $\theta_j = 0.328\pi$. Furthermore, within the angular range $\theta_j \in [0.242, 0.408]\pi$, the homodyne-detection sensitivity surpasses the SQL, demonstrating a clear quantum advantage.
 
 Next, we analyze the configuration with a coherent state combined with a squeezed vacuum state as input. This means $\hat{a}_{\mathrm{in}}$ is coherent state $|\alpha\rangle$ with $\alpha = |\alpha|e^{i\theta_\alpha}$. While $\hat{b}_{\mathrm{in}}$ is squeezed vacuum state $\hat{S}(\xi)|0\rangle$ with $\xi=re^{i\theta_r}$. The parameters are set as $g = 0.7$, $\theta_{\alpha} = \theta_r = 0$, $\theta_{\beta} = \pi$, and we choose $|\alpha| = 10^7$ to ensure that the two input configurations have the same total input photon number.
 
 Fig.~\ref{fig:5}(b) displays the corresponding sensitivity for squeezing parameters $r=0$ (green dashed line), $r=1$ (magenta dotted line), and $r=2$ (blue dashed line), together with the SQL (red solid line) and QCRB (black solid line) for comparison. It can be observed that appropriately increasing the squeeze parameter $r$ enhances the sensitivity. Over most of the angular domain away from $0$ and $\pi/2$, the sensitivity $\Delta\theta_j$ also remains below \SI{e-3}{\degree}. The optimal sensitivity of \SI{1.20e-6}{\degree} is achieved around $\theta_j = 0.316\pi$ for $r=2$. For $r=0$ (corresponding to a vacuum input in mode $\hat{b}_{\mathrm{in}}$), the sensitivity is comparable to the dual-coherent-state case and surpasses the SQL only over a relatively narrow interval $\theta_j \in [0.222\pi,\, 0.418\pi]$. In contrast, for $r=1,2$ (squeezed vacuum input), the sensitivity surpasses the SQL over a substantially broader range, $\theta_j \in [0,\, 0.423\pi]$.
 
 We emphasize that the two polarization angles, $\theta_a$ and $\theta_p$, are physically equivalent. Although their determination procedures differ (as illustrated in Fig.~\ref{fig:4}), they represent the same fundamental degree of freedom of the field polarization. Consequently, our calculations yield identical sensitivities for both angles. Moreover, the angular region over which our scheme surpasses the SQL is not fixed but can be tuned dynamically. By adjusting the magnetic field $B$ and the coupling laser detuning $\Delta_c$, we can selectively bring a chosen transition pathway (e.g., transition 1 in Fig.~\ref{fig:2}(c)) near resonance. This enhances the coupling to a specific spherical field component (e.g., $\boldsymbol{\epsilon}_0$), thereby creating a tunable window of high sensitivity beyond the SQL centered at a desired polarization angle $\theta_j$. This dynamic tunability enables the sensitive region to be tailored for specific practical applications.

\subsection{Sensitivity optimization}\label{sec:subsection_V_C}
So far, we have analyzed how the sensitivity depends on the polarization angles $\theta_j$ for different input states. We now turn to the influence of other key parameters, which not only affect the sensitivity but also determine the uniqueness of the polarization angle extraction.

\begin{figure}[tbp]
	\label{Fig6} \centering \includegraphics[width=1\columnwidth]{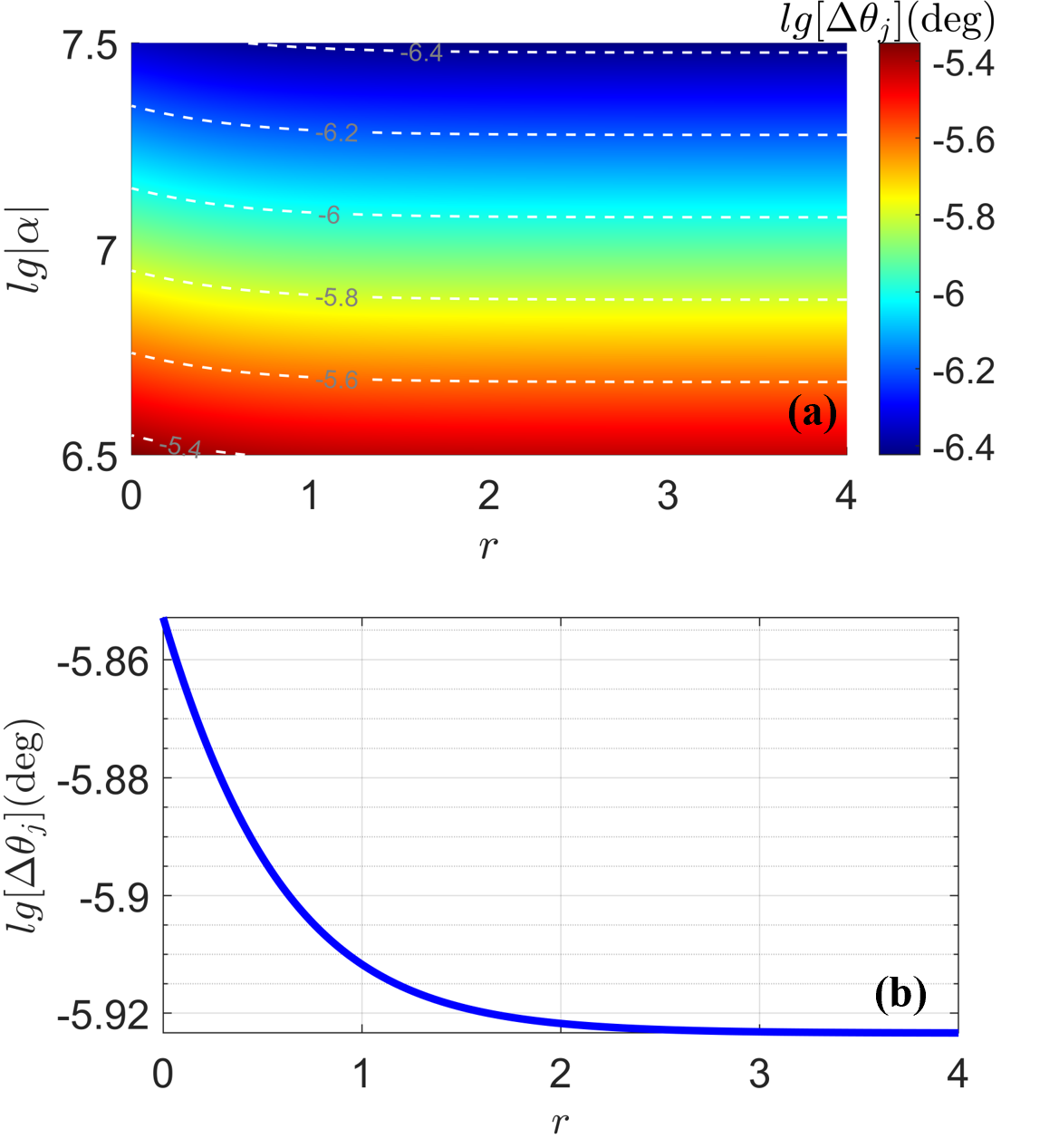}%
	\newline
	\caption{(a) Optimal polarization angle sensitivity $\Delta\theta_j$ as a function of the injected coherence amplitude $|\alpha|$ and the squeeze amplitude $r$. The white dash line is the contour line. (b) $\Delta\theta_j$ as a function of squeeze amplitude $r$ with $|\alpha|=10^7$. Other parameters are chosen to be the same as in Fig.~\ref{fig:4}
	}\label{fig:6}
\end{figure}

We first examine the dependence of the optimal polarization angle sensitivity on the coherent amplitude $|\alpha|$ and the squeezing amplitude $r$. Figure~\ref{fig:6} shows this optimal sensitivity as a function of both parameters.

It is clearly seen that increasing the coherent amplitude $|\alpha|$ improves the sensitivity. However, higher $|\alpha|$ also increases collision rates and power broadening. Balancing these competing effects leads to a practical choice of $|\alpha| = 10^7$, as used in our previous calculations. Similarly, a larger squeezing amplitude $r$ enhances the sensitivity, but the improvement saturates around $r \approx 2$; beyond this point, further increasing $r$ yields diminishing returns.

\begin{figure}[tbp]
	\centering
	\includegraphics[width=1\columnwidth]{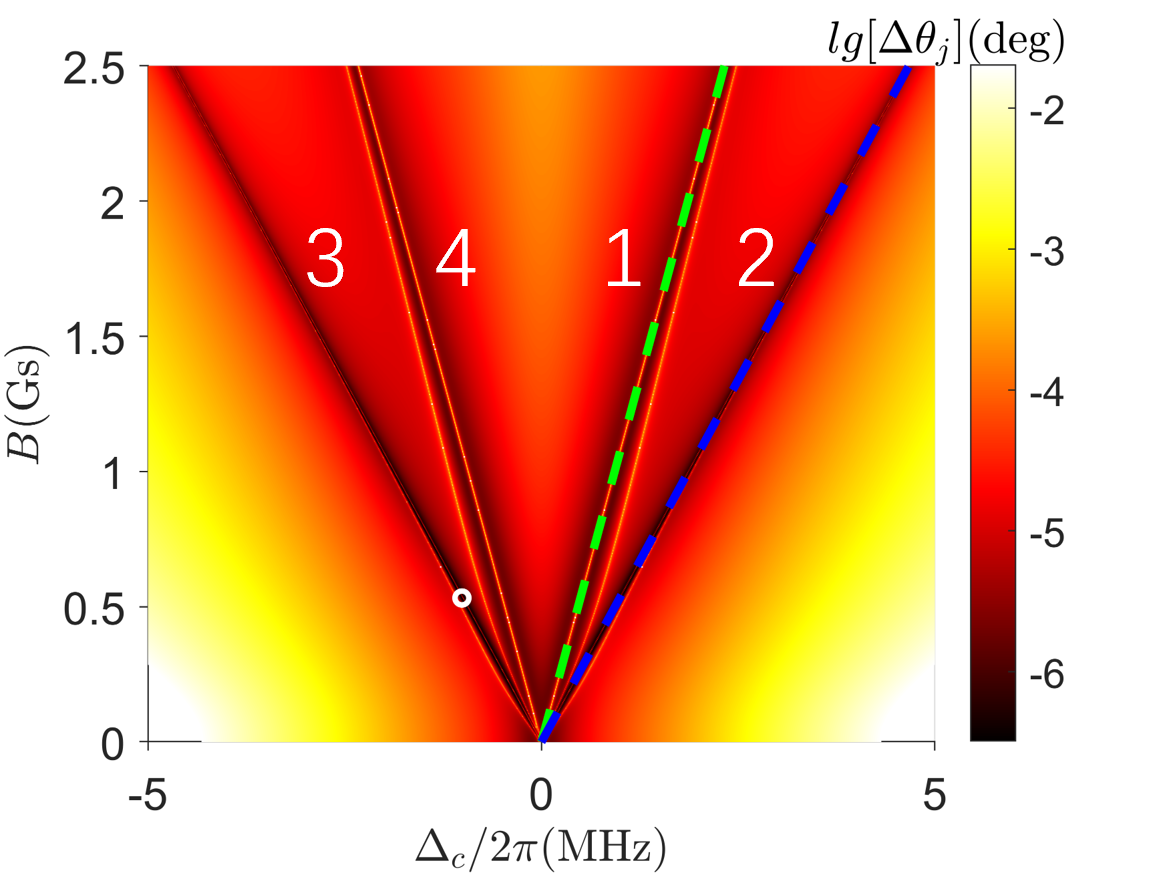}
	\caption{
		Optimal sensitivity $\Delta\theta_j$ as a function of magnetic field strength $B$ and coupling laser detuning $\Delta_c$ for dual coherent-state inputs. The green dashed line traces the parameters $(B, \Delta_c)$ for which the RF field is resonantly coupled to transition 1 (see Fig.~\ref{fig:2}(c)); the blue dashed line corresponds to resonant coupling with transition 2. The parameter sets for resonant coupling with transitions 3 and 4 are symmetric to these two lines on the plot and are omitted for clarity. White open circles mark the locations of optimal sensitivity.
	}
	\label{fig:7}
\end{figure}

As noted in Sec.~\ref{sec:subsection_V_B}, the magnetic field strength $B$ and the coupling laser detuning $\Delta_c$ strongly affect the angular region where the sensitivity surpasses the SQL. We now examine their direct influence on the achievable sensitivity. Figure~\ref{fig:7} presents the optimal sensitivity $\Delta\theta_j$ as a function of $B$ and $\Delta_c$ for the dual-coherent-state input configuration.As noted in Sec.~\ref{sec:subsection_V_B}, the magnetic field strength $B$ and the coupling laser detuning $\Delta_c$ significantly influence the angular region where the sensitivity surpasses the SQL (see Sec.~V.B). We now investigate their direct impact on the sensitivity itself. The optimal sensitivity $\Delta\theta_j$ as a function of $B$ and $\Delta_c$ with dual coherent states as input is presented in Fig.~\ref{fig:7}.

As illustrated in Fig.~\ref{fig:7}, the green and blue dashed lines trace the parameters $(B, \Delta_c)$ for which the RF field is resonantly coupled to transitions 1 and 2, respectively; the symmetric lines for transitions 3 and 4 are omitted for clarity. The figure reveals that relatively high sensitivity (low $\Delta\theta_j$) is achieved when $(B, \Delta_c)$ lies near any of the four resonant-coupling lines. Particularly low sensitivities are obtained when the system is near-resonantly coupled to transition 2 or 3. The global sensitivity minimum (marked by the white open circle) occurs at $B = 0.56\,\mathrm{G}$ and $\Delta_c = -2\pi\,\mathrm{MHz}$, which are the parameters adopted in Sec.~\ref{sec:subsection_V_A}. The symmetric point $(B, \Delta_c) = (0.56\,\mathrm{G}, +2\pi\,\mathrm{MHz})$ yields an identical minimum but is not marked.

\section{CONCLUSION}\label{sec:section_VI}
	In conclusion, we have presented a theoretical framework for high-precision vector polarimetry that combines Rydberg atoms with an SU(1,1) interferometer. Here the probe mode after OPA1 of the SU(1,1) interferometer couples to the ground-state transition of the Rydberg atoms, while the RF electric field is coupled to the Rydberg-state transition. A weak static magnetic field lifts the Zeeman degeneracy, creating four distinct transition pathways whose differential coupling to the spherical polarization components of the RF field encodes the polarization angles into the atomic absorption index. By solving the Heisenberg–Langevin equations, we derived the susceptibility $\chi$ of the atomic medium, which we modeled as a phase change $\theta$ followed by a fictitious beam splitter of transmittance $e^{-\epsilon}$. This model was integrated into the SU(1,1) interferometer, and the resulting input–output relation allowed the atomic absorption index $\epsilon$ to be retrieved with high sensitivity via homodyne detection.
	
	We subsequently calculated the sensitivity of polarization angle measurements via homodyne detection and obtained the corresponding SQL and the QCRB. For the case of dual coherent states as input, homodyne detection enables sensitivity surpass the SQL within an angular range of $\sim0.16\pi$. For a coherent state combined with a squeezed vacuum state  as input, the sensitivity surpasses the SQL starting from $\theta_j = 0$ over a substantially broader range of about $0.42\pi$. Furthermore, we analyzed the influence of key parameters---including the squeezing parameter, injected coherent amplitude, magnetic field strength, and detuning of the coupling light---on the sensitivity, identifying optimal operating conditions.
	
	Overall, our scheme enables unique resolution of both polarization angles $\theta_a$ and $\theta_p$ across the full range $[0, \pi/2]$. The sensitivity remains below \SI{e-3}{\degree} over most of this range, and can reach below \SI{e-6}{\degree} under optimal conditions. The approach is particularly suited to weak-field detection, offers flexible parameter tuning to place the surpass-SQL sensitivity window in a desired angular region, and enables independent measurement of field amplitude and polarization.
	
	Our work establishes a high-precision vector-polarimetry scheme that exploits quantum resources to significantly enhance sensitivity. The methodology can be extended to measure other electromagnetic field parameters or to DOA estimation. It holds considerable potential for applications in Rydberg‑atom‑based quantum sensing, quantum communication, and quantum information processing.
	
	\section{ACKNOWLEDGEMENTS}\label{sec:section_VII}
	This work was supported by Natural Science Foundation of Shaanxi Province (Grant No. 2024JC-YBMS-031); Shaanxi Fundamental Science Research Project of Mathematics and Physics (Grant No. 22JSY005); National Natural Science Foundation of China (11534008, 91536115);
	
	\bibliographystyle{unsrt} 
	\bibliography{myrefs}     
	
\end{document}